\documentclass[12pt]{article}

\usepackage{geometry}
\usepackage{amsmath}
\usepackage{mathrsfs}  
\usepackage{mathtools} 
\usepackage{bbold}    
\usepackage{braket}    
\usepackage{gensymb}   

\usepackage{graphicx}      	
\usepackage{subcaption}
\usepackage{multirow}       
\usepackage{here} 			
\usepackage{xurl}		
\usepackage[plainpages=false,pdfpagelabels=true, linkcolor=cyan,breaklinks,colorlinks=true]{hyperref}  
\usepackage{soul}                   
\usepackage[dvipsnames]{xcolor}		
\usepackage{siunitx}
\usepackage{xspace}
\usepackage{booktabs}    
\usepackage{authblk}	
\usepackage[numbers]{natbib} 
\usepackage{etoolbox} 

\newcommand{\flo}[1]{{\color{Green}#1}}

\newcommand{\ute}{UTe\textsubscript{2}\xspace}
\newcommand{\cij}[1]{\ensuremath{c_{#1}}\xspace}
\newcommand{\aij}[1]{\ensuremath{\alpha_{#1}}\xspace}
\newcommand{\cvs}{CsV\textsubscript{3}Sb\textsubscript{5}\xspace}

\newcommand{\Tc}{\ensuremath{T_{\rm c}}\xspace}

\newcommand{\Tcdw}{\ensuremath{T_{\rm CDW}}\xspace}

\begin{document}

\title{Absence of a Bulk Thermodynamic Phase Transition to a Density Wave Phase in UTe$_2$}%

\author[1]{Florian Theuss}%
\author[1]{Avi Shragai}
\author[1,2,3]{Gael Grissonnanche}
\author[4]{Luciano Peralta}%
\author[1]{Gregorio de la Fuente Simarro}
\author[5]{Ian M Hayes}
\author[5]{Shanta R Saha}
\author[5]{Yun Suk Eo}
\author[5]{Alonso Suarez}
\author[6]{Andrea Capa Salinas}
\author[6]{Ganesh Pokharel}
\author[6]{Stephen D. Wilson}
\author[5,7]{Nicholas P Butch}
\author[5,8]{Johnpierre Paglione}
\author[1,8]{B.~J.~Ramshaw \thanks{bradramshaw@cornell.edu}}

\affil[1]{Laboratory of Atomic and Solid State Physics, Cornell University, Ithaca, NY 14853, USA}
\affil[2]{Kavli Institute at Cornell for Nanoscale Science, Ithaca, New York, USA}
\affil[3]{Laboratoire des Solides Irradiés, CEA/DRF/lRAMIS, CNRS, École Polytechnique, Institut Polytechnique de Paris, F-91128 Palaiseau, France}
\affil[4]{Department of Physics, Universidad de Los Andes, Bogot´a, 111711, Colombia}
\affil[5]{Maryland Quantum Materials Center, Department of Physics, University of Maryland, College Park, Maryland 20742, USA}
\affil[6]{Materials Department and California Nanosystems Institute, University of California Santa Barbara, Santa Barbara, CA, USA}
\affil[7]{NIST Center for Neutron Research, National Institute of Standards and Technology, 100 Bureau Drive, Gaithersburg, Maryland 20899, USA}
\affil[8]{Canadian Institute for Advanced Research, Toronto, Ontario, Canada}

\newpage
\date{}%
\maketitle

\newpage
\section*{Abstract}
\textbf{
Competing and intertwined orders are ubiquitous in strongly correlated electron systems, such as the charge, spin, and superconducting orders in the high-$\boldsymbol{T_{\rm c}}$ cuprates. Recent scanning tunneling microscopy (STM) measurements provide evidence for a charge density wave (CDW) that coexists with superconductivity in the heavy Fermion metal \ute. This CDW persists up to at least 7.5 K and, as a CDW breaks the translational symmetry of the lattice, its disappearance is necessarily accompanied by thermodynamic phase transition. Here, we report high-precision thermodynamic measurements of the elastic moduli of \ute. We observe no signature of a phase transition in the elastic moduli down to a level of 1 part in $\boldsymbol{10^7}$, strongly implying the absence of bulk CDW order in \ute. We suggest that the CDW and associated pair density wave (PDW) observed by STM may be confined to the surface of \ute. }

\section*{Introduction}

The superconductivity of \ute is unconventional in many respects: it is spin triplet \cite{nakamine2019superconducting}, it has a re-entrant phase at very high magnetic fields \cite{ran2019nearly, ran_extreme_2019}, and its superconducting \Tc bifurcates into two transitions under pressure \cite{braithwaiteMultipleSuperconductingPhases2019,aokiMultipleSuperconductingPhases2020}. Recent STM experiments \cite{aishwarya2023MagneticfieldsensitiveChargeDensity,lafleur2023inhomogeneous,gu2023DetectionPairDensity,aishwarya2024MeltingChargeDensity} have provided evidence for even more unusual behavior: the coexistence of superconductivity with an incommensurate CDW. Upon applying a magnetic field, superconductivity and the CDW are suppressed at the same critical field \cite{aishwarya2023MagneticfieldsensitiveChargeDensity}. This observation suggests a close connection between superconductivity and the CDW in \ute, potentially connected via a parent PDW \cite{gu2023DetectionPairDensity}.

STM measurements clearly indicate that the CDW persists up to at least 7.5 K---more than a factor of three higher than \Tc---and disappears by a temperature no higher than 12~K \cite{lafleur2023inhomogeneous}. A corollary of this observation is that, upon cooling below 12~K, \ute first enters a broken-symmetry CDW phase, followed by the onset of superconductivity at lower temperature. As broken-symmetry phases are necessarily accompanied by thermodynamic phase transitions \cite{landau1937theory}, it is natural to ask whether such a phase transition is observed between 7.5 and 12~K in \ute. The answer thus far is negative: the superconducting transition is the only phase transition visible in specific heat measurements at ambient pressures and in zero magnetic field \cite{willa2021thermodynamic}. 

The apparent absence of a second phase transition leads to the following question: does CDW order exist in the bulk of \ute, or is it confined to the surface? This gets to the heart of a broader problem: which of the many exotic phenomena that have been discovered in \ute are representative of the bulk, and which are particular to the surface? For example, polar Kerr effect \cite{hayes2021multicomponent}, STM \cite{jiao2020chiral}, and microwave conductivity \cite{bae2021anomalous} measurements all suggest a two-component, time-reversal symmetry breaking order parameter, whereas ultrasound \cite{theuss2024single} and some specific heat \cite{rosa2022single} measurements suggest a single-component order parameter. One possible resolution is that Kerr, STM, and microwaves are all sensitive to a unique superconducting state on the surface of \ute, whereas ultrasound and specific heat are sensitive to the bulk order parameter. This issue of bulk versus surface superconductivity in UTe$_2$ is closely related to the existence of a CDW, as a superconducting PDW is necessarily accompanied by a CDW. The existence of such a PDW/CDW pair in the bulk would strongly constrain the microscopic mechanism of Cooper pairing in \ute.

To investigate the possibility of a phase transition to bulk CDW order in \ute, we measure the elastic moduli as a function of temperature, from 2~K to 280~K. Elastic moduli are particularly sensitive to CDW phase transitions because they break the translational symmetry of the lattice. This has been investigated extensively in other CDW systems, such as the rare-earth tritellurides, where discontinuities in the elastic moduli at \Tcdw are of order a few times $10^{-2}$ of the total elastic moduli. Because elastic moduli can be measured with better than one part in $10^{7}$ precision, our measurements have many decades of sensitivity to explore the possibility of a bulk phase transition to CDW order in \ute.

\section*{Results}

Elastic moduli, $c_{ij}$, are the thermodynamic coefficients characterizing the susceptibility of a material to strain. In terms of the total free energy $\mathcal{F}$, elastic moduli are given by
\begin{equation}
	c_{ij} = \frac{\partial^2 \mathcal{F}}{\partial \epsilon_{ij}^2},
	\label{eq:moduli}
\end{equation}
where $\epsilon_{ij}$ is a particular component of the strain tensor. These moduli are related to the sound velocities, $v_{ij}$, by $v_{ij} = \sqrt{c_{ij}/\rho}$, where $\rho$ is the material density. Like other thermodynamic susceptibilities, such as specific heat, elastic moduli exhibit singular behaviour at phase transitions \cite{landau2013statistical}. 

\begin{figure}[H]
	\begin{center}
	\includegraphics[width=1\textwidth,trim=0.2 0 0.7cm 0.1cm,clip]{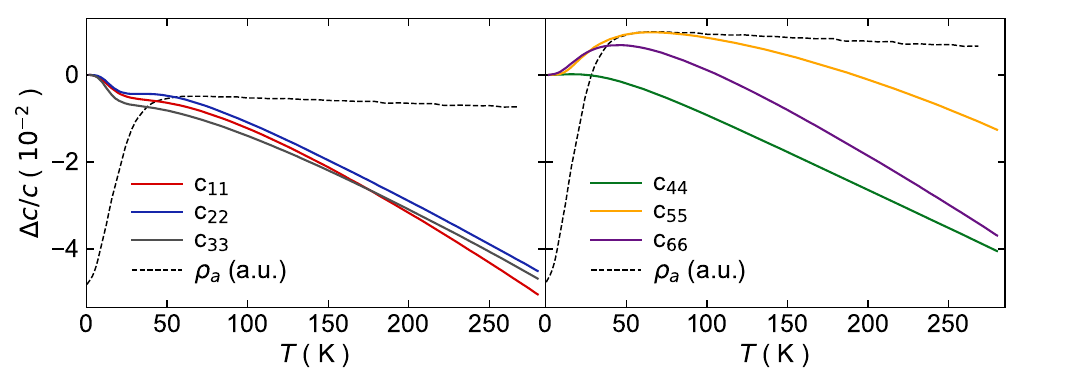}
	\end{center}
	\caption{ \textbf{Elastic moduli of \ute from 2 K to 280 K.} The three compressional moduli (left) and shear moduli (right) are shown from 1.2 to 300 kelvin. $\Delta c/c$ is defined  as $(c(T)-c(2~ {\rm K}))/c(2~ {\rm K})$.  The $a$-axis resistivity of \ute is plotted as a dashed line for comparison and is taken from \citet{PhysRevB.106.L060505}. The downturn of the resistivity around 50 kelvin coincides with deviations from conventional stiffening due to lattice anharmonicity in $c_{11}$, $c_{22}$, $c_{33}$, $c_{55}$, and $c_{66}$.}
	\vspace{-0em}
	\label{fig:pulse_broad}
\end{figure}

We first show the elastic moduli corresponding to all six unique strains in \ute measured over a broad temperature range---from 2 K to 280 K---using pulse echo ultrasound. This data is reproduced from \citet{theuss2024single}, where details about the sample preparation and experimental technique are given. 

All six elastic moduli exhibit only smooth behaviour across the entire temperature range down to \Tc. The shear modulus $c_{44}$ exhibits conventional stiffening due to lattice anharmonicity \cite{varshni1970temperature}. The other two shear ($c_{55}$ and $c_{66}$) and three compressional moduli ($c_{11}$, $c_{22}$, and $c_{33}$) exhibit smooth evolution with temperature that is associated with the onset of Kondo coherence near 50 kelvin \flo{\cite{luthi1985magnetoacoustics,luthi1994ElectronPhononEffects}}. This demonstrates the sensitivity of 5 out of 6 elastic moduli to changes in the electronic structure of \ute. Despite this sensitivity, no sharp changes in slope or discontinuities that would be indicative of an electronic phase transition are visible on this scale.

\begin{figure}[H]
	\begin{center}
		\includegraphics[width=1\textwidth,trim=0.2 0 0.9cm 0.1cm,clip]{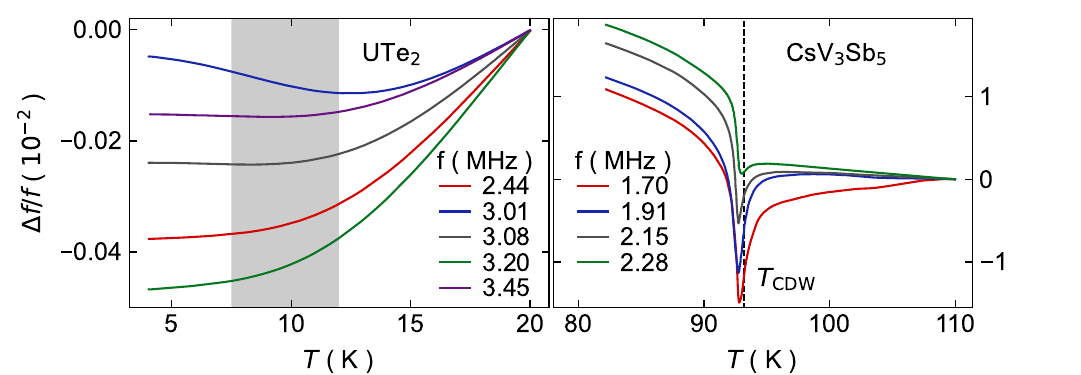}
	\end{center}
	\caption{ \textbf{Low temperature elastic resonances of \cvs and \ute .} High resolution mechanical resonances of \ute (left) and \cvs (right). Each curve tracks the frequency of a single mechanical resonance of a single-crystal sample. $\Delta f/f$ is defined as $(f(T)-f(T_0))/f(T_0)$, where $T_0$ is the highest temperature shown. A grey bar indicates the temperature range over which the CDW disappears in STM measurements of \ute.  }
	\vspace{-0em}
	\label{fig:CVS_comp}
\end{figure}

To further constrain the presence or absence of a CDW transition, we performed high-resolution resonant ultrasound spectroscopy (RUS) measurements on \ute across the temperature range where the CDW peaks disappear in the STM experiments \cite{lafleur2023inhomogeneous}. RUS measures the mechanical resonance frequencies of a sample. Each resonance frequency is determined by the sample geometry, the material density, and the elastic moduli. While procedures exist for decomposing the temperature dependence of the resonance frequencies into the temperature dependence of the elastic moduli \cite{visscher1991normal,ramshaw2015avoided,theuss2024resonant} (see Methods), the highest signal to noise is obtained by directly examining the resonance frequencies.

\autoref{fig:CVS_comp} shows the temperature dependence of 5 resonance frequencies from 4 to 20 K---across the temperature range where the CDW disappears in the STM experiments. No anomalies are visible across the entire temperature range. These resonance frequencies contain admixtures of all nine elastic moduli in different proportions: the top-most resonance (blue) is dominated by compressional moduli, whereas the resonance at the bottom of the figure (green) is dominated by shear moduli. Thus, all resonances should show a singular jump, as well as a change in slope, at a phase transition \cite{rehwald1973study}. Using the signal-to-noise of our measurement, we can constrain any singularity in the elastic moduli to be smaller than $1\times10^{-7}$ (see Methods).

To provide a sense of scale for what is expected at a CDW transition, we perform similar RUS measurements on \cvs. \cvs has both a CDW transition near 90 K and a superconducting \Tc near 2 K \cite{ortiz2020cs}. The analogy with \ute is close, as there is also STM evidence for a PDW in this material \cite{chen2021roton}. The impact of the CDW phase transition on the elastic moduli is striking: discontinuties on the order of $2\times10^{-2}$ are visible at \Tcdw---five orders of magnitude larger than our signal to noise in \ute. Similar-sized anomalies at \Tcdw are present in the elastic moduli of the rare earth tritellurides, and anomalies of order $10^{-4}$ are found at the transition to the high-field CDW phase of the high-\Tc cuprates. These results are summarized, along with other examples from the literature, in \autoref{tab:dc/c jumps comparison}.

\begin{table}
	\centering
	\begin{tabular}{c c c c c}
		Material  & $T_{CDW}$ ( K ) & $\Delta c/c$ & $c_{ij}$ & Reference \\
		\toprule
		\ute  & $10-12$            & $<1\times10^{-7}$         &  & This work\\
		\cvs  & 93            & $\approx 10^{-2}$         &  & This work\\
		\midrule
		TbTe$_3$  & 333            & $2\times 10^{-2}$         & $c_{11}$& \citet{saint-paul2016ElasticAnomaliesCharge}\\
		& & $2\times 10^{-2}$ & $c_{33}$ &\\
		K$_{0.3}$MoO$_3$  & 180            & $2\times 10^{-2}$         & $Y_\mathrm{[102]}$ & \citet{brill1995ThermodynamicsChargeDensityWaveTransition}\\
		Lu$_5$Ir$_4$Si$_{10}$ & 80 & $1\times 10^{-2}$ & $c_{11}$& \citet{saint-paul2020ElasticAnomaliesFirst} \\
		& & $6\times 10^{-3}$ &  $c_{33}$ &\\
		TTF-TCNQ  & 53            & $1\times 10^{-2}$         & $Y_\mathrm{[010]}$& \citet{barmatz1974elastic}\\
		2$H$-NbSe$_2$ & 30 & $1\times 10^{-3}$ & $Y_\mathrm{[100]}$ & \citet{barmatz1975ElasticityMeasurementsLayered}\\
		YBa$_2$Cu$_3$O$_{6.55}$ ($H=30$~T) & 50            & $8\times 10^{-5}$         & $c_{22}$ & \citet{laliberte2018HighFieldCharge}\\
	
		\bottomrule
	\end{tabular}
	\caption{\textbf{Elastic moduli discontinuities in several CDW materials.} $Y_\mathrm{[ijk]}$ is a Youngs modulus measured with the compressive stress in the $\mathrm{[ijk]}$ direction. NbSe$_2$, \cvs, YBa$_2$Cu$_3$O$_{6.55}$, and Lu$_5$Ir$_4$Si$_{10}$ also have superconducting transitions at lower temperatures. The elastic anomaly in TbTe$_3$ is representative of all rare-earth tritellurides \cite{saint2021phenomenological}.}
	\label{tab:dc/c jumps comparison}
\end{table}

\section*{Discussion}

Our data constrain any thermodynamic signature of a CDW phase transition in \ute to be $\Delta c/c < 1\times 10^{-7}$. This is five orders of magnitude smaller than what is observed in \cvs and in the rare-earth tritellurides \cite{saint-paul2016ElasticAnomaliesCharge}, and three orders of magnitude smaller than what is observed in the high-\Tc cuprates. There are two common arguments as to why a CDW might not exhibit a thermodynamic signature in the elastic moduli: from broadening of the transition due to disorder, and from insensitivity of the CDW to the lattice (a ``purely electronic" CDW). We address these possibilities in turn.

Disorder tends to broaden all thermodynamic singularities at a CDW transition \cite{singh2005competition,luccas2015charge}. This is most apparent in the high-\Tc cuprates. There, CDW correlation lengths range from a few unit cells in Bi$_2$Sr$_2$CaCu$_2$O$_{8 + \delta}$ \cite{hoffman2002four}, to roughly 100 angstroms in YBa$_2$Cu$_3$O$_{6+\delta}$ \cite{wu2011magnetic,ghiringhelli2012long,blanco2014resonant}. Because of the short correlation length, no anomaly is found in either the specific heat or the elastic moduli at the onset of the CDW correlations \cite{loram2001evidence,laliberte2018HighFieldCharge}. Only upon applying a magnetic field is the correlation length increased to roughly 300 angstroms and an associated singularity observed in the elastic moduli \cite{gerber2015three,laliberte2018HighFieldCharge}. The CDW seen by STM in \ute is qualitatively different from that found in the cuprates: in \ute, the CDW peaks are of a similar width to the crystalline Bragg peaks \cite{aishwarya2023MagneticfieldsensitiveChargeDensity}. Furthermore, existing ultrasound investigations into the superconducting state of \ute reveal sharp superconducting transitions---with widths only 5~\% of \Tc---that show no signs of broadening due to disorder \cite{theuss2024single}. Thus disorder is an unlikely explanation for the lack of thermodynamic singularity at a putative \Tcdw.


A ``purely electronic" CDW would be a CDW that is entirely decoupled from the crystalline lattice. To our knowledge, such a state does not (and likely cannot) exist. The very fact that \ute has a non-Galilean-invariant (i.e.~non-spherical) Fermi surface \cite{aoki2022first} means that the electronic degrees of freedom are coupled to the lattice potential. As sound waves deform that lattice potential, they are necessarily coupled to the conduction electrons. This is evidenced by softening of the elastic moduli at the onset of Kondo coherence near 50 K (\autoref{fig:pulse_broad}), as well as by the sharp thermodynamic singularities seen at \Tc \cite{theuss2024single}. Any redistribution of the charge density in a metal is necessarily compensated for by a displacement of the lattice to maintain local charge neutrality, and thus a thermodynamic signature in the elastic moduli at a CDW transition is inescapable.

We find no evidence for a transition to a CDW phase in the bulk of \ute, despite such evidence existing on the (011) surface as measured by STM \cite{aishwarya2023MagneticfieldsensitiveChargeDensity,lafleur2023inhomogeneous,gu2023DetectionPairDensity,aishwarya2024MeltingChargeDensity}. This suggests that the surface of \ute may host different ordered states than the bulk, including possibly a different superconducting order parameter. Thus, while the host of unconventional phenomena  discovered by both bulk and surface measurements in \ute must ultimately be understood within a single framework, one should be careful when extrapolating observations made on the surface to the bulk and vice versa.

\section*{Acknowledgments}
A. Shragai, B. J. R., and F. T. acknowledge funding from the Office of Basic Energy Sciences of the United States Department of Energy under award no. DE-SC0020143 (ultrasound experiments and analysis). N. B. and J. P. acknowledge support from the Department of Energy award number DE-SC-0019154 (sample characterization), the Gordon and Betty Moore Foundation’s EPiQS Initiative through grant number GBMF9071 (materials synthesis), the National Science Foundation under grant number DMR-2105191 (sample preparation), and the Maryland Quantum Materials Center and the National Institute of Standards and Technology. B. J. R. and F. T. acknowledge use of the Cornell Center for Materials Research Shared Facilities which are supported through the NSF MRSEC program (DMR-1719875).

\section*{Author Contributions}
B. J. R. conceived the experiment.
I. M. H, S. R. S, Y. S. E, A. Suarez, A. C. S, and G. P grew and characterized the samples.
F. T. and A. Shragai performed the sample preparation and transducer fabrication.
F. T., A. Shragai, and G.G. performed the ultrasound measurements.
F. T., and B. J. R. performed the data analysis.
F. T. and B. J. R. wrote the manuscript with input from all other co-authors.
S. D. W., J. P., N. P. B., and B. J. R. supervised the project.

\section*{Competing Interests}
The authors declare no competing interests.


\begin{thebibliography}{45}
	\providecommand{\natexlab}[1]{#1}
	\providecommand{\url}[1]{\texttt{#1}}
	\expandafter\ifx\csname urlstyle\endcsname\relax
	\providecommand{\doi}[1]{doi: #1}\else
	\providecommand{\doi}{doi: \begingroup \urlstyle{rm}\Url}\fi
	
	\bibitem[Aishwarya et~al.(2023)Aishwarya, {May-Mann}, Raghavan, Nie, Romanelli,
	Ran, Saha, Paglione, Butch, Fradkin, and
	Madhavan]{aishwarya2023MagneticfieldsensitiveChargeDensity}
	Anuva Aishwarya, Julian {May-Mann}, Arjun Raghavan, Laimei Nie, Marisa
	Romanelli, Sheng Ran, Shanta~R. Saha, Johnpierre Paglione, Nicholas~P. Butch,
	Eduardo Fradkin, and Vidya Madhavan.
	\newblock Magnetic-field-sensitive charge density waves in the superconductor
	{UTe$_2$}.
	\newblock \emph{Nature}, 618\penalty0 (7967):\penalty0 928--933, June 2023.
	\newblock ISSN 1476-4687.
	\newblock \doi{10.1038/s41586-023-06005-8}.
	
	\bibitem[Aishwarya et~al.(2024)Aishwarya, {May-Mann}, Almoalem, Ran, Saha,
	Paglione, Butch, Fradkin, and Madhavan]{aishwarya2024MeltingChargeDensity}
	Anuva Aishwarya, Julian {May-Mann}, Avior Almoalem, Sheng Ran, Shanta~R. Saha,
	Johnpierre Paglione, Nicholas~P. Butch, Eduardo Fradkin, and Vidya Madhavan.
	\newblock Melting of the charge density wave by generation of pairs of
	topological defects in {UTe$_2$}.
	\newblock \emph{Nature Physics}, March 2024.
	\newblock ISSN 1745-2481.
	\newblock \doi{10.1038/s41567-024-02429-9}.
	
	\bibitem[Aoki et~al.(2020)Aoki, Honda, Knebel, Braithwaite, Nakamura, Li,
	Homma, Shimizu, Sato, Brison, and
	Flouquet]{aokiMultipleSuperconductingPhases2020}
	Dai Aoki, Fuminori Honda, Georg Knebel, Daniel Braithwaite, Ai~Nakamura, De~Xin
	Li, Yoshiya Homma, Yusei Shimizu, Yoshiki~J. Sato, Jean~Pascal Brison, and
	Jacques Flouquet.
	\newblock Multiple superconducting phases and unusual enhancement of the upper
	critical field in {UTe$_2$}.
	\newblock \emph{Journal of the Physical Society of Japan}, 89\penalty0
	(5):\penalty0 053705, May 2020.
	\newblock ISSN 13474073.
	\newblock \doi{10.7566/JPSJ.89.053705}.
	
	\bibitem[Aoki et~al.(2022)Aoki, Sakai, Opletal, Tokiwa, Ishizuka, Yanase,
	Harima, Nakamura, Li, Homma, et~al.]{aoki2022first}
	Dai Aoki, Hironori Sakai, Petr Opletal, Yoshifumi Tokiwa, Jun Ishizuka, Youichi
	Yanase, Hisatomo Harima, Ai~Nakamura, Dexin Li, Yoshiya Homma, et~al.
	\newblock First observation of the de haas--van alphen effect and fermi
	surfaces in the unconventional superconductor {UTe$_2$}.
	\newblock \emph{Journal of the Physical Society of Japan}, 91\penalty0
	(8):\penalty0 083704, 2022.
	
	\bibitem[Bae et~al.(2021)Bae, Kim, Eo, Ran, Liu, Fuhrman, Paglione, Butch, and
	Anlage]{bae2021anomalous}
	Seokjin Bae, Hyunsoo Kim, Yun~Suk Eo, Sheng Ran, I-lin Liu, Wesley~T Fuhrman,
	Johnpierre Paglione, Nicholas~P Butch, and Steven~M Anlage.
	\newblock Anomalous normal fluid response in a chiral superconductor {UTe$_2$}.
	\newblock \emph{Nature communications}, 12\penalty0 (1):\penalty0 2644, 2021.
	
	\bibitem[Barmatz et~al.(1974)Barmatz, Testardi, Garito, and
	Heeger]{barmatz1974elastic}
	M~Barmatz, LR~Testardi, AF~Garito, and AJ~Heeger.
	\newblock Elastic properties of one dimensional compounds.
	\newblock \emph{Solid State Communications}, 15\penalty0 (8):\penalty0
	1299--1302, 1974.
	
	\bibitem[Barmatz et~al.(1975)Barmatz, Testardi, and
	Di~Salvo]{barmatz1975ElasticityMeasurementsLayered}
	M.~Barmatz, L.~R. Testardi, and F.~J. Di~Salvo.
	\newblock Elasticity measurements in the layered dichalcogenides {TaSe$_2$} and
	{NbSe$_2$}.
	\newblock \emph{Physical Review B}, 12\penalty0 (10):\penalty0 4367--4376,
	November 1975.
	\newblock \doi{10.1103/PhysRevB.12.4367}.
	
	\bibitem[Blanco-Canosa et~al.(2014)Blanco-Canosa, Frano, Schierle, Porras,
	Loew, Minola, Bluschke, Weschke, Keimer, and Le~Tacon]{blanco2014resonant}
	S.~Blanco-Canosa, A.~Frano, E.~Schierle, J.~Porras, T.~Loew, M.~Minola,
	M.~Bluschke, E.~Weschke, B.~Keimer, and M.~Le~Tacon.
	\newblock Resonant x-ray scattering study of charge-density wave correlations
	in {YBa$_2$Cu$_3$O$_{6+x}$}.
	\newblock \emph{Phys. Rev. B}, 90:\penalty0 054513, Aug 2014.
	\newblock \doi{10.1103/PhysRevB.90.054513}.
	\newblock URL \url{https://link.aps.org/doi/10.1103/PhysRevB.90.054513}.
	
	\bibitem[Braithwaite et~al.(2019)Braithwaite, Vali{\v s}ka, Knebel, Lapertot,
	Brison, Pourret, Zhitomirsky, Flouquet, Honda, and
	Aoki]{braithwaiteMultipleSuperconductingPhases2019}
	D.~Braithwaite, M.~Vali{\v s}ka, G.~Knebel, G.~Lapertot, J.-P. Brison,
	A.~Pourret, M.~E. Zhitomirsky, J.~Flouquet, F.~Honda, and D.~Aoki.
	\newblock Multiple superconducting phases in a nearly ferromagnetic system.
	\newblock \emph{Communications Physics}, 2\penalty0 (1):\penalty0 147, December
	2019.
	\newblock ISSN 2399-3650.
	\newblock \doi{10.1038/s42005-019-0248-z}.
	
	\bibitem[Brill et~al.(1995)Brill, Chung, Kuo, Zhan, Figueroa, and
	Mozurkewich]{brill1995ThermodynamicsChargeDensityWaveTransition}
	J.~W. Brill, M.~Chung, Y.~K. Kuo, X.~Zhan, E.~Figueroa, and George Mozurkewich.
	\newblock Thermodynamics of the charge-density-wave transition in blue bronze.
	\newblock \emph{Physical Review Letters}, 74\penalty0 (7):\penalty0 1182--1185,
	February 1995.
	\newblock \doi{10.1103/PhysRevLett.74.1182}.
	
	\bibitem[Chen et~al.(2021)Chen, Yang, Hu, Zhao, Yuan, Xing, Qian, Huang, Li,
	Ye, et~al.]{chen2021roton}
	Hui Chen, Haitao Yang, Bin Hu, Zhen Zhao, Jie Yuan, Yuqing Xing, Guojian Qian,
	Zihao Huang, Geng Li, Yuhan Ye, et~al.
	\newblock Roton pair density wave in a strong-coupling kagome superconductor.
	\newblock \emph{Nature}, 599\penalty0 (7884):\penalty0 222--228, 2021.
	
	\bibitem[Eo et~al.(2022)Eo, Liu, Saha, Kim, Ran, Horn, Hodovanets, Collini,
	Metz, Fuhrman, Nevidomskyy, Denlinger, Butch, Fuhrer, Wray, and
	Paglione]{PhysRevB.106.L060505}
	Yun~Suk Eo, Shouzheng Liu, Shanta~R. Saha, Hyunsoo Kim, Sheng Ran, Jarryd~A.
	Horn, Halyna Hodovanets, John Collini, Tristin Metz, Wesley~T. Fuhrman,
	Andriy~H. Nevidomskyy, Jonathan~D. Denlinger, Nicholas~P. Butch, Michael~S.
	Fuhrer, L.~Andrew Wray, and Johnpierre Paglione.
	\newblock $c$-axis transport in {UTe$_2$}: Evidence of three-dimensional
	conductivity component.
	\newblock \emph{Phys. Rev. B}, 106:\penalty0 L060505, Aug 2022.
	\newblock \doi{10.1103/PhysRevB.106.L060505}.
	\newblock URL \url{https://link.aps.org/doi/10.1103/PhysRevB.106.L060505}.
	
	\bibitem[Gerber et~al.(2015)Gerber, Jang, Nojiri, Matsuzawa, Yasumura, Bonn,
	Liang, Hardy, Islam, Mehta, et~al.]{gerber2015three}
	Simon Gerber, H~Jang, H~Nojiri, S~Matsuzawa, H~Yasumura, DA~Bonn, R~Liang,
	WN~Hardy, Z~Islam, A~Mehta, et~al.
	\newblock Three-dimensional charge density wave order in
	{YBa$_2$Cu$_3$O$_{6.67}$} at high magnetic fields.
	\newblock \emph{Science}, 350\penalty0 (6263):\penalty0 949--952, 2015.
	
	\bibitem[Ghiringhelli et~al.(2012)Ghiringhelli, Le~Tacon, Minola,
	Blanco-Canosa, Mazzoli, Brookes, De~Luca, Frano, Hawthorn, He,
	et~al.]{ghiringhelli2012long}
	G~Ghiringhelli, M~Le~Tacon, Matteo Minola, S~Blanco-Canosa, Claudio Mazzoli,
	NB~Brookes, GM~De~Luca, A~Frano, DG~Hawthorn, F~He, et~al.
	\newblock Long-range incommensurate charge fluctuations in {(Y, Nd)}
	{Ba$_2$Cu$_3$O$_{6+ x}$}.
	\newblock \emph{Science}, 337\penalty0 (6096):\penalty0 821--825, 2012.
	
	\bibitem[Ghosh et~al.(2021)Ghosh, Shekhter, Jerzembeck, Kikugawa, Sokolov,
	Brando, Mackenzie, Hicks, and Ramshaw]{ghosh2021thermodynamic}
	Sayak Ghosh, Arkady Shekhter, F~Jerzembeck, N~Kikugawa, Dmitry~A Sokolov,
	Manuel Brando, AP~Mackenzie, Clifford~W Hicks, and BJ~Ramshaw.
	\newblock Thermodynamic evidence for a two-component superconducting order
	parameter in {Sr$_2$RuO$_4$}.
	\newblock \emph{Nature Physics}, 17\penalty0 (2):\penalty0 199--204, 2021.
	
	\bibitem[Gu et~al.(2023)Gu, Carroll, Wang, Ran, Broyles, Siddiquee, Butch,
	Saha, Paglione, Davis, and Liu]{gu2023DetectionPairDensity}
	Qiangqiang Gu, Joseph~P. Carroll, Shuqiu Wang, Sheng Ran, Christopher Broyles,
	Hasan Siddiquee, Nicholas~P. Butch, Shanta~R. Saha, Johnpierre Paglione,
	J.~C.~S{\'e}amus Davis, and Xiaolong Liu.
	\newblock Detection of a pair density wave state in {UTe$_2$}.
	\newblock \emph{Nature}, 618\penalty0 (7967):\penalty0 921--927, June 2023.
	\newblock ISSN 1476-4687.
	\newblock \doi{10.1038/s41586-023-05919-7}.
	
	\bibitem[Hayes et~al.(2021)Hayes, Wei, Metz, Zhang, Eo, Ran, Saha, Collini,
	Butch, Agterberg, et~al.]{hayes2021multicomponent}
	Ian~M Hayes, Di~S Wei, Tristin Metz, Jian Zhang, Yun~Suk Eo, Sheng Ran,
	Shanta~R Saha, John Collini, Nicholas~P Butch, Daniel~F Agterberg, et~al.
	\newblock Multicomponent superconducting order parameter in ute2.
	\newblock \emph{Science}, 373\penalty0 (6556):\penalty0 797--801, 2021.
	
	\bibitem[Hoffman et~al.(2002)Hoffman, Hudson, Lang, Madhavan, Eisaki, Uchida,
	and Davis]{hoffman2002four}
	Jennifer~E Hoffman, Eric~W Hudson, KM~Lang, Vidya Madhavan, Hiroshi Eisaki,
	Shin’ichi Uchida, and James~C Davis.
	\newblock A four unit cell periodic pattern of quasi-particle states
	surrounding vortex cores in {Bi$_2$Sr$_2$CaCu$_2$O$_{8+\delta}$}.
	\newblock \emph{Science}, 295\penalty0 (5554):\penalty0 466--469, 2002.
	
	\bibitem[Jiao et~al.(2020)Jiao, Howard, Ran, Wang, Rodriguez, Sigrist, Wang,
	Butch, and Madhavan]{jiao2020chiral}
	Lin Jiao, Sean Howard, Sheng Ran, Zhenyu Wang, Jorge~Olivares Rodriguez,
	Manfred Sigrist, Ziqiang Wang, Nicholas~P Butch, and Vidya Madhavan.
	\newblock Chiral superconductivity in heavy-fermion metal {UTe$_2$}.
	\newblock \emph{Nature}, 579\penalty0 (7800):\penalty0 523--527, 2020.
	
	\bibitem[LaFleur et~al.(2024)LaFleur, Li, Frank, Xu, Cheng, Wang, Butch, and
	Zeljkovic]{lafleur2023inhomogeneous}
	Alexander LaFleur, Hong Li, Corey~E. Frank, Muxian Xu, Siyu Cheng, Ziqiang
	Wang, Nicholas~P. Butch, and Ilija Zeljkovic.
	\newblock Inhomogeneous high temperature melting and decoupling of charge
	density waves in spin-triplet superconductor {UTe$_2$}.
	\newblock \emph{Nature Communications}, 15:\penalty0 4456, 2024.
	\newblock \doi{https://doi.org/10.1038/s41467-024-48844-7}.
	
	\bibitem[Lalibert{\'e} et~al.(2018)Lalibert{\'e}, Frachet, Benhabib, Borgnic,
	Loew, Porras, Le~Tacon, Keimer, Wiedmann, Proust, and
	LeBoeuf]{laliberte2018HighFieldCharge}
	Francis Lalibert{\'e}, Mehdi Frachet, Siham Benhabib, Benjamin Borgnic,
	Toshinao Loew, Juan Porras, Mathieu Le~Tacon, Bernhard Keimer, Steffen
	Wiedmann, Cyril Proust, and David LeBoeuf.
	\newblock High field charge order across the phase diagram of
	{YBa$_2$Cu$_3$O$_y$}.
	\newblock \emph{npj Quantum Materials}, 3\penalty0 (1):\penalty0 1--7, March
	2018.
	\newblock ISSN 2397-4648.
	\newblock \doi{10.1038/s41535-018-0084-5}.
	
	\bibitem[Landau(1937)]{landau1937theory}
	Lev~Davidovich Landau.
	\newblock On the theory of phase transitions. i.
	\newblock \emph{Zh. Eksp. Teor. Fiz.}, 11:\penalty0 19, 1937.
	
	\bibitem[Landau and Lifshitz(2013)]{landau2013statistical}
	Lev~Davidovich Landau and Evgenii~Mikhailovich Lifshitz.
	\newblock \emph{Statistical Physics: Volume 5}, volume~5.
	\newblock Elsevier, 2013.
	
	\bibitem[Loram et~al.(2001)Loram, Luo, Cooper, Liang, and
	Tallon]{loram2001evidence}
	JW~Loram, J~Luo, JR~Cooper, WY~Liang, and JL~Tallon.
	\newblock Evidence on the pseudogap and condensate from the electronic specific
	heat.
	\newblock \emph{Journal of Physics and Chemistry of Solids}, 62\penalty0
	(1-2):\penalty0 59--64, 2001.
	
	\bibitem[Luccas et~al.(2015)Luccas, Fente, Hanko, Correa-Orellana, Herrera,
	Climent-Pascual, Azpeitia, P\'erez-Casta\~neda, Osorio, Salas-Colera, Nemes,
	Mompean, Garc\'{\i}a-Hern\'andez, Rodrigo, Ramos, Guillam\'on, Vieira, and
	Suderow]{luccas2015charge}
	R.~F. Luccas, A.~Fente, J.~Hanko, A.~Correa-Orellana, E.~Herrera,
	E.~Climent-Pascual, J.~Azpeitia, T.~P\'erez-Casta\~neda, M.~R. Osorio,
	E.~Salas-Colera, N.~M. Nemes, F.~J. Mompean, M.~Garc\'{\i}a-Hern\'andez,
	J.~G. Rodrigo, M.~A. Ramos, I.~Guillam\'on, S.~Vieira, and H.~Suderow.
	\newblock Charge density wave in layered {La$_{1-x}$Ce$_x$Sb$_2$}.
	\newblock \emph{Phys. Rev. B}, 92:\penalty0 235153, Dec 2015.
	\newblock \doi{10.1103/PhysRevB.92.235153}.
	\newblock URL \url{https://link.aps.org/doi/10.1103/PhysRevB.92.235153}.
	
	\bibitem[L{\"u}thi(1985)]{luthi1985magnetoacoustics}
	B~L{\"u}thi.
	\newblock Magnetoacoustics in intermetallic f-electron systems.
	\newblock \emph{Journal of magnetism and magnetic materials}, 52\penalty0
	(1-4):\penalty0 70--78, 1985.
	
	\bibitem[L{\"u}thi et~al.(1994)L{\"u}thi, Bruls, Thalmeier, Wolf, Finsterbusch,
	and Kouroudis]{luthi1994ElectronPhononEffects}
	B.~L{\"u}thi, G.~Bruls, P.~Thalmeier, B.~Wolf, D.~Finsterbusch, and
	I.~Kouroudis.
	\newblock Electron-phonon effects in heavy fermion systems.
	\newblock \emph{Journal of Low Temperature Physics}, 95\penalty0 (1):\penalty0
	257--270, April 1994.
	\newblock ISSN 1573-7357.
	\newblock \doi{10.1007/BF00754941}.
	
	\bibitem[Nakamine et~al.(2019)Nakamine, Kitagawa, Ishida, Tokunaga, Sakai,
	Kambe, Nakamura, Shimizu, Homma, Li, et~al.]{nakamine2019superconducting}
	Genki Nakamine, Shunsaku Kitagawa, Kenji Ishida, Yo~Tokunaga, Hironori Sakai,
	Shinsaku Kambe, Ai~Nakamura, Yusei Shimizu, Yoshiya Homma, Dexin Li, et~al.
	\newblock Superconducting properties of heavy {Fermion UTe$_2$} revealed by
	{$^125$Te}-nuclear magnetic resonance.
	\newblock \emph{journal of the physical society of japan}, 88\penalty0
	(11):\penalty0 113703, 2019.
	
	\bibitem[Ortiz et~al.(2020)Ortiz, Teicher, Hu, Zuo, Sarte, Schueller, Abeykoon,
	Krogstad, Rosenkranz, Osborn, et~al.]{ortiz2020cs}
	Brenden~R Ortiz, Samuel~ML Teicher, Yong Hu, Julia~L Zuo, Paul~M Sarte, Emily~C
	Schueller, AM~Milinda Abeykoon, Matthew~J Krogstad, Stephan Rosenkranz,
	Raymond Osborn, et~al.
	\newblock Cs v 3 sb 5: A z 2 topological kagome metal with a superconducting
	ground state.
	\newblock \emph{Physical Review Letters}, 125\penalty0 (24):\penalty0 247002,
	2020.
	
	\bibitem[Ramshaw et~al.(2015)Ramshaw, Shekhter, McDonald, Betts, Mitchell,
	Tobash, Mielke, Bauer, and Migliori]{ramshaw2015avoided}
	BJ~Ramshaw, Arkady Shekhter, Ross~D McDonald, Jon~B Betts, JN~Mitchell,
	PH~Tobash, CH~Mielke, ED~Bauer, and Albert Migliori.
	\newblock Avoided valence transition in a plutonium superconductor.
	\newblock \emph{Proceedings of the National Academy of Sciences}, 112\penalty0
	(11):\penalty0 3285--3289, 2015.
	
	\bibitem[Ran et~al.(2019{\natexlab{a}})Ran, Eckberg, Ding, Furukawa, Metz,
	Saha, Liu, Zic, Kim, Paglione, et~al.]{ran2019nearly}
	Sheng Ran, Chris Eckberg, Qing-Ping Ding, Yuji Furukawa, Tristin Metz, Shanta~R
	Saha, I-Lin Liu, Mark Zic, Hyunsoo Kim, Johnpierre Paglione, et~al.
	\newblock Nearly ferromagnetic spin-triplet superconductivity.
	\newblock \emph{Science}, 365\penalty0 (6454):\penalty0 684--687,
	2019{\natexlab{a}}.
	
	\bibitem[Ran et~al.(2019{\natexlab{b}})Ran, Liu, Eo, Campbell, Neves, Fuhrman,
	Saha, Eckberg, Kim, Graf, Balakirev, Singleton, Paglione, and
	Butch]{ran_extreme_2019}
	Sheng Ran, I.-Lin Liu, Yun~Suk Eo, Daniel~J. Campbell, Paul~M. Neves, Wesley~T.
	Fuhrman, Shanta~R. Saha, Christopher Eckberg, Hyunsoo Kim, David Graf, Fedor
	Balakirev, John Singleton, Johnpierre Paglione, and Nicholas~P. Butch.
	\newblock Extreme magnetic field-boosted superconductivity.
	\newblock \emph{Nature Physics}, 15\penalty0 (12):\penalty0 1250--1254,
	December 2019{\natexlab{b}}.
	\newblock ISSN 1745-2481.
	\newblock \doi{10.1038/s41567-019-0670-x}.
	
	\bibitem[Ran et~al.(2021)Ran, Liu, Saha, Saraf, Paglione, and
	Butch]{ran2021comparison}
	Sheng Ran, I-Lin Liu, Shanta~R Saha, Prathum Saraf, Johnpierre Paglione, and
	Nicholas~P Butch.
	\newblock Comparison of two different synthesis methods of single crystals of
	superconducting uranium ditelluride.
	\newblock \emph{JoVE (Journal of Visualized Experiments)}, \penalty0
	(173):\penalty0 e62563, 2021.
	
	\bibitem[Rehwald(1973)]{rehwald1973study}
	Walther Rehwald.
	\newblock The study of structural phase transitions by means of ultrasonic
	experiments.
	\newblock \emph{Advances in Physics}, 22\penalty0 (6):\penalty0 721--755, 1973.
	
	\bibitem[Rosa et~al.(2022)Rosa, Weiland, Fender, Scott, Ronning, Thompson,
	Bauer, and Thomas]{rosa2022single}
	Priscila~FS Rosa, Ashley Weiland, Shannon~S Fender, Brian~L Scott, Filip
	Ronning, Joe~D Thompson, Eric~D Bauer, and Sean~M Thomas.
	\newblock Single thermodynamic transition at 2 k in superconducting ute2 single
	crystals.
	\newblock \emph{Communications Materials}, 3\penalty0 (1):\penalty0 33, 2022.
	
	\bibitem[Saint-Paul and Monceau(2021)]{saint2021phenomenological}
	M~Saint-Paul and P~Monceau.
	\newblock Phenomenological approach of the thermodynamic properties of the
	charge density wave systems.
	\newblock \emph{Philosophical Magazine}, 101\penalty0 (5):\penalty0 598--621,
	2021.
	
	\bibitem[{Saint-Paul} et~al.(2016){Saint-Paul}, Guttin, Lejay, Remenyi,
	Leynaud, and Monceau]{saint-paul2016ElasticAnomaliesCharge}
	M.~{Saint-Paul}, C.~Guttin, P.~Lejay, G.~Remenyi, O.~Leynaud, and P.~Monceau.
	\newblock Elastic anomalies at the charge density wave transition in
	{TbTe$_3$}.
	\newblock \emph{Solid State Communications}, 233:\penalty0 24--29, May 2016.
	\newblock ISSN 0038-1098.
	\newblock \doi{10.1016/j.ssc.2016.02.008}.
	
	\bibitem[{Saint-Paul} et~al.(2020){Saint-Paul}, Opagiste, and
	Guttin]{saint-paul2020ElasticAnomaliesFirst}
	M.~{Saint-Paul}, C.~Opagiste, and C.~Guttin.
	\newblock Elastic anomalies at the first order transition in
	{Lu$_5$Ir$_4$Si$_{10}$}.
	\newblock \emph{Journal of Physics and Chemistry of Solids}, 138:\penalty0
	109255, March 2020.
	\newblock ISSN 0022-3697.
	\newblock \doi{10.1016/j.jpcs.2019.109255}.
	
	\bibitem[Singh et~al.(2005)Singh, Nirmala, Ramakrishnan, and
	Malik]{singh2005competition}
	Yogesh Singh, R.~Nirmala, S.~Ramakrishnan, and S.~K. Malik.
	\newblock Competition between superconductivity and charge-density-wave
	ordering in the {Lu$_5$Ir$_4$(Si$_{1-x}$Ge$_x$)$_{10}$} alloy system.
	\newblock \emph{Phys. Rev. B}, 72:\penalty0 045106, Jul 2005.
	\newblock \doi{10.1103/PhysRevB.72.045106}.
	\newblock URL \url{https://link.aps.org/doi/10.1103/PhysRevB.72.045106}.
	
	\bibitem[Theuss et~al.(2024{\natexlab{a}})Theuss, Shragai, Grissonnanche,
	Hayes, Saha, Eo, Suarez, Shishidou, Butch, Paglione,
	et~al.]{theuss2024single}
	Florian Theuss, Avi Shragai, Gael Grissonnanche, Ian~M Hayes, Shanta~R Saha,
	Yun~Suk Eo, Alonso Suarez, Tatsuya Shishidou, Nicholas~P Butch, Johnpierre
	Paglione, et~al.
	\newblock Single-component superconductivity in {UTe$_2$} at ambient pressure.
	\newblock \emph{Nature Physics}, pages 1--7, 2024{\natexlab{a}}.
	
	\bibitem[Theuss et~al.(2024{\natexlab{b}})Theuss, Simarro, Shragai,
	Grissonnanche, Hayes, Saha, Shishidou, Chen, Nakatsuji, Ran,
	et~al.]{theuss2024resonant}
	Florian Theuss, Gregorio de la~Fuente Simarro, Avi Shragai, Gael Grissonnanche,
	Ian~M Hayes, Shanta Saha, Tatsuya Shishidou, Taishi Chen, Satoru Nakatsuji,
	Sheng Ran, et~al.
	\newblock Resonant ultrasound spectroscopy for irregularly shaped samples and
	its application to uranium ditelluride.
	\newblock \emph{Physical Review Letters}, 132\penalty0 (6):\penalty0 066003,
	2024{\natexlab{b}}.
	
	\bibitem[Varshni(1970)]{varshni1970temperature}
	YP~Varshni.
	\newblock Temperature dependence of the elastic constants.
	\newblock \emph{Physical Review B}, 2\penalty0 (10):\penalty0 3952, 1970.
	
	\bibitem[Visscher et~al.(1991)Visscher, Migliori, Bell, and
	Reinert]{visscher1991normal}
	William~M Visscher, Albert Migliori, Thomas~M Bell, and Robert~A Reinert.
	\newblock On the normal modes of free vibration of inhomogeneous and
	anisotropic elastic objects.
	\newblock \emph{The Journal of the Acoustical Society of America}, 90\penalty0
	(4):\penalty0 2154--2162, 1991.
	
	\bibitem[Willa et~al.(2021)Willa, Hardy, Aoki, Li, Wiecki, Lapertot, and
	Meingast]{willa2021thermodynamic}
	Kristin Willa, Fr\'ed\'eric Hardy, Dai Aoki, Dexin Li, Paul Wiecki, G\'erard
	Lapertot, and Christoph Meingast.
	\newblock Thermodynamic signatures of short-range magnetic correlations in
	{UTe$_{2}$}.
	\newblock \emph{Phys. Rev. B}, 104:\penalty0 205107, Nov 2021.
	\newblock \doi{10.1103/PhysRevB.104.205107}.
	\newblock URL \url{https://link.aps.org/doi/10.1103/PhysRevB.104.205107}.
	
	\bibitem[Wu et~al.(2011)Wu, Mayaffre, Kr{\"a}mer, Horvati{\'c}, Berthier,
	Hardy, Liang, Bonn, and Julien]{wu2011magnetic}
	Tao Wu, Hadrien Mayaffre, Steffen Kr{\"a}mer, Mladen Horvati{\'c}, Claude
	Berthier, WN~Hardy, Ruixing Liang, DA~Bonn, and Marc-Henri Julien.
	\newblock Magnetic-field-induced charge-stripe order in the high-temperature
	superconductor {YBa$_2$Cu$_3$O$_y$}.
	\newblock \emph{Nature}, 477\penalty0 (7363):\penalty0 191--194, 2011.
	
\end{thebibliography}

\section*{Methods}

\subsection*{Sample preparation}
\subsubsection*{\cvs}
Single crystals of \cvs were synthesized by the self-flux method in an inert environment. Elemental liquid Cs (Alfa 99.98\%), V powder (Sigma 99.9\%) in-house pre-purified in a 9:1 ethanol/hydrochloric acid mixture, and Sb shot (Alfa 99.999\%) were weighed out to Cs$_{20}$V$_{15}$Sb$_{120}$ stoichiometry and milled in a tungsten carbide vial. The precursor milled powder was heated up to 1000 °C, soaked for 12 hours, cooled down to 900 °C at 5 °C/h and further cooled down to 500 °C at 2 °C. Once at room temperature, the resulting crystals were extracted manually in air. 

A single-crystal specimen was selected for RUS based on visual inspection. Samples with cracks, excessive flux, and intergrowth of secondary phases were avoided due to their detrimental effects on the mechanical quality factors.

\subsubsection*{\ute}

Single crystals of \ute were grown by the chemical vapour transport method as described in \citet{ran2021comparison}. 

The surface of a single-crystal specimen was digitized using a Zeiss Xradia Versa XRM-520 X-ray nano-CT, and the digitized mesh was aligned to the crystal axes using back-reflection Laue. 

\subsection*{Experiments}
\subsubsection*{Pulse-echo ultrasound}
Measurements were performed in an Oxford Instruments Heliox $^3$He refrigerator using a traditional phase-comparison pulse-echo method. Measurements were performed on 3 different samples in 6 different transducer configurations. More details about the samples and experiment can be found in the Methods of \citet{theuss2024single}.

\subsubsection*{Resonant ultrasound spectroscopy}

We performed RUS on a single-crystal sample of \ute in a custom-built probe immersed in a bath of $^4$He. Details of the apparatus and the procedure for measuring resonance frequencies can be found in the methods of \citet{ghosh2021thermodynamic}. Temperature sweeps were performed using a slow ramp rate of approximately 0.025 K/min. 

We fit the elastic moduli of this sample at 4 K using the method described in \citet{theuss2024resonant}. The absolute elastic moduli for this sample are found in Table III of \citet{theuss2024resonant} under Sample B. The raw resonance frequencies, as well as the fit we obtain, are found in table XI of the supplementary to \citet{theuss2024resonant}. 

The fit to the resonance spectrum allows us to determine the contribution from each elastic modulus to each resonance frequency. Each resonance frequency $f_k$ contains contributions from all elastic moduli, and the relative change in resonance frequency as a function of temperature is given by 
\begin{equation}
	\frac{\Delta f_k}{f_k}  =  \sum_{i,j} \alpha_{ij}^{(k)} \frac{\Delta c_{ij}}{c_{ij}},
	\label{eq:alpha coefficients}
\end{equation}
where the $\alpha_{ij}^{(k)}$ coefficients are temperature independent, $\Delta f_k/f_k \equiv (f_k(T)-f_k(T_0))/f_k(T_0)$ is the relative change in resonance frequency referenced to temperature $T_0$, and likewise for $\Delta c_{ij}/c_{ij}$. The coefficients $\alpha_{ij}^{(k)}$ sum to one for each resonance: $\sum_{i,j} \alpha_{ij}^{(k)} = 1$.

\autoref{fig:CVS_comp} shows 5 resonance frequencies selected for having high $Q$ factors ($>10^5$) and qualitatively different temperature dependencies such that all elastic moduli are represented in this data set. This representation is quantified through the $\alpha_{ij}^{(k)}$ coefficients of \autoref{eq:alpha coefficients}. \autoref{tab:freqs} shows these coefficients for the 5 resonance frequencies plotted in \autoref{fig:CVS_comp}. 

Discontinuities are expected in the three compressional elastic moduli---\cij{11}, \cij{22}, and \cij{33}---at any phase transition \cite{ghosh2021thermodynamic}, irrespective of which symmetries are broken (shear moduli can have discontinuities for some, but not all order parameters). Thus, it is critical that the three compressional moduli are well-represented in the resonance frequencies we analyze. \autoref{tab:freqs} shows that the temperature dependence of \cij{11}, \cij{22}, and \cij{33} each make up between 10\% and 30\% of the total frequency shift for the 5 resonances shown in \autoref{fig:CVS_comp}. Thus if there was a phase transition, these 5 frequencies would show it. 

\begin{table}
	\centering
	\begin{tabular}{c | c c c c c c c c c }
		$f$ (MHz)  & \aij{11} & \aij{22}& \aij{33}& \aij{12}& \aij{13}& \aij{23}& \aij{44}& \aij{55}& \aij{66} \\
		\toprule
		2.44		& 0.141	&	0.088 &	0.212 &	0.004  & -0.092 & -0.028 & 0.263 & 0.129 & 0.283 \\
		3.01		& 0.250	&	0.178 &	0.310 & -0.022 & -0.152 & -0.045 & 0.190 & 0.168 & 0.123 \\
		3.08		& 0.241 &   0.113 & 0.190 & -0.030 & -0.115 & -0.007 & 0.233 & 0.129 & 0.246 \\
		3.20		& 0.191 &   0.082 & 0.244 & -0.004 & -0.134 & -0.020 & 0.246 & 0.146 & 0.249 \\
		3.45		& 0.212 &   0.108 & 0.190 & -0.015 & -0.109 & -0.018 & 0.248 & 0.110 & 0.274 \\
		\bottomrule
	\end{tabular}
	\caption{\textbf{Resonance frequency composition.} The coefficient of each elastic modulus that makes up the resonance frequency as defined in \autoref{eq:alpha coefficients}.}
	\label{tab:freqs}
\end{table}

\subsection*{Noise analysis}

\begin{figure}[H]
	\begin{center}
		\includegraphics[width=.5\textwidth]{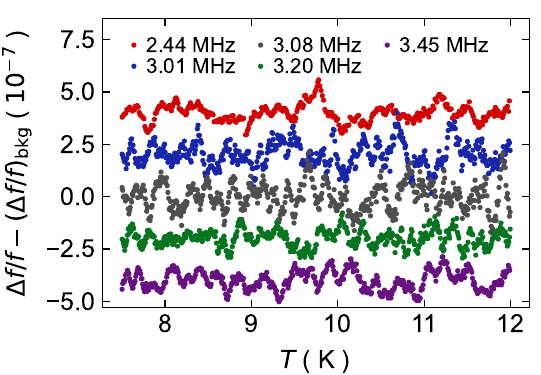}
	\end{center}
	\caption{ \textbf{Noise level of \ute resonances.} The same data as presented in \autoref{fig:CVS_comp} but with a $5^{\rm th}$ order polynomial background subtracted. The spacing between points is approximately 8 mK, and the data have had a moving average applied over a 10 point window. The data have been offset vertically for clarity.}
	\vspace{-0em}
	\label{fig:noise analysis}
\end{figure}

\autoref{fig:noise analysis} shows the same \ute RUS data from \autoref{fig:CVS_comp}, but with a $5^{\rm th}$ order polynomial subtracted and a moving average of 10 points applied. With a temperature step of approximately 8 mK, this averages over an 80 mK window---the same width as the thermodynamic singularity at \Tc in \ute \cite{theuss2024single}. The noise on this scale is of order $\Delta f/f \approx \pm 1\times10^{-7}$.

\end{document}